\documentclass[aps,pre,twocolumn,groupedaddress,showpacs]{revtex4}
\usepackage{graphicx}
\usepackage{dcolumn}
\usepackage{bm}
\usepackage{epsfig}
\usepackage{color}
\usepackage{float}
\input epsf
\epsfclipon

\begin{document}
\title{Finite size induces crossover temperature in growing spin chains} 
\author{Julian Sienkiewicz, Krzysztof Suchecki, and Janusz A. Ho{\l}yst}
\affiliation{Faculty of Physics, Center of Excellence for Complex Systems Research, Warsaw University of Technology, Koszykowa 75, PL-00-662 Warsaw, Poland}
\date{\today}
\begin{abstract}
We introduce a growing one-dimensional quenched spin model that bases on  asymmetrical one-side Ising interactions in the presence of external field. Numerical simulations and analytical calculations based on Markov chain theory show that when the external field is smaller than the exchange coupling constant $J$ there is a non-monotonous dependence of the mean magnetization on the temperature in a finite system. The crossover temperature $T_c$ corresponding to the maximal magnetization decays with system size, approximately as the inverse of the W Lambert function. The observed phenomenon can be understood as an interplay between the thermal fluctuations and the presence of the first cluster determined by initial conditions. The effect exists also when  spins are not quenched but fully thermalized after the attachment to the chain. We conceive the model is suitable for a qualitative description of online emotional discussions arranged in a chronological order, where a spin in every node conveys emotional valence of a subsequent post.
\end{abstract} 
\maketitle

\section{INTRODUCTION}

Due to their simplicity and fully analytical treatment, one-dimensional models are useful and comprehensible objects for theoretical studies. Of the exceptional importance backed by the feasibility of calculations is the Ising model \cite{ising, ruelle, dyson, frohlich, thouless, grinstein, bak, denisov, yilmaz, percus, chiara, koffel}. Such a system with short-range ferromagnetic interactions possesses no crossover temperature when system's susceptibility is observed. This is true for a non-growing system and when each spin is symmetrically coupled to its left and right neighbor \cite{statmech}. In this paper we introduce an evolving spin model with an asymmetrical one-side dynamics. However, the asymmetry is unlike the one proposed by Huang \cite{huang, chakraborty}, where the spin variable can take on two eigenvalues +1 and $-1/\lambda$ with $\lambda > 1$ nor it is connected to the degeneration of higher-energy spin state \cite{mansfield}. Instead, we explicitly modify Ising Hamiltonian by taking into account only node's left neighbor as well as equip our model with a growing component (a new node is quenched after a single update). We show, numerically and analytically, that it results in a {\it crossover temperature} correspoding to the maximal susceptibility when the system is finite and the field is smaller than the spin interaction constant. This unexpected phenomenon is further explained as an interplay between the thermal fluctuations and the first spin cluster determined by initial conditions.

Although one-dimensional systems are frequently used to model social dynamics \cite{sznajd, rumor, kondratiuk, isolation, isolation2}, such an approach often suffers from over-simplicity, e.g., one finds no evidence to support the idea that agents related to social interactions are to be distributed on a chain. In this paper we give clear reasons for choosing this very topology. In fact, our model is motivated by the recent research \cite{plos, ania} on affective interactions among participants of  Internet fora \cite{frank2, frank, bosa}. Such media often use a chronological structure of the incoming posts that can be easily regarded as a one-dimensional chain (i.e., the consecutive posts are represented by the nodes in the chain). The results of our previous analyses \cite{plos,ania} indicate that one of the most dominant phenomena seen in such media is a strong dependence of the expressed emotion on the emotion of the last comment (i.e., the newest one).

\section{MODEL DESCRIPTION}
\begin{figure}
\centering
\includegraphics[width=0.7\columnwidth]{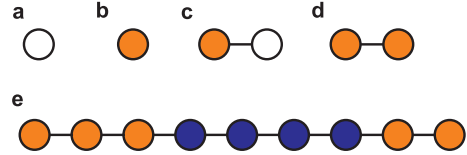}
\caption{(color online) A scheme of the simulation process. (a-d) Consecutive steps of an exemplary simulation: (a) starting from an empty node, (b) adding random spin to the first node, (c) adding the next node, (d) inserting spin according to dynamics rule. (e) An effect of the simulation. Orange ($s=+1$) and blue ($s=-1$) discs symbolize spins.}
\label{fig:eis_scheme}
\end{figure}

The model bases on the idea of a growing chain (see Fig. \ref{fig:eis_scheme}). The process is organized as follows: the first node of the chain has a random spin $s_0=\pm 1$ (that could be interpreted as emotional valence \cite{valence} of a post in online discussion), that is drawn with probability $\Pr(s_0= \pm 1)=1/2$ (Fig. \ref{fig:eis_scheme}a-b). Then, another node of the chain is added to the right side of the last one (Fig. \ref{fig:eis_scheme}c) and it is initially equipped with a spin once again drawn with equal probabilities $\Pr(s_1= \pm 1) = 1/2$. Subsequently, the node becomes a subject to the updating procedure that is based on the Ising-like model approach (Fig. \ref{fig:eis_scheme}d). For each new node $n$ we define a function $\mathcal{E}_n = - J s_{n-1} s_n - h s_n$, where the constant $J > 0$ corresponds to exchange integral in the Ising model and $h$ is the external field. A minimum of the function $\mathcal{E}_n$ conforms to spins of the same sign in the consecutive nodes of the chain, thus $\mathcal{E}_n$ can be treated as an {\it emotional discomfort} function felt by a user posting a message $s_n$. As the spin is drawn, we test how flipping its sign to the opposite one (i.e., from $s_n=+1$ to $s_n=-1$ or likewise) affects the change of function $\mathcal{E}$ as $\Delta \mathcal{E} = \mathcal{E}'_n - \mathcal{E}_n = - (Js_{n-1} + h)(s'_n-s_n)$, where term $\mathcal{E}'_n$ corresponds to $s_n'$ calculated when $s_n \rightarrow s'_n =-s_n$. Then we follow the Metropolis algorithm \cite{metropolis} i.e., if the $\Delta \mathcal{E}< 0$ we accept the change, otherwise we test if the expression $\exp[-\Delta \mathcal{E}(k_B T)^{-1}]$ is smaller or larger than a random value $\xi \in [0; 1]$ (here $k_B$ is Boltzmann constant and $T$ is temperature). If the latter occurs we accept the change, otherwise the spin is kept as originally chosen. The procedure of adding new nodes and setting their spin variables according to the above described rules is repeated until the size $N$ of the chain is reached (Fig. \ref{fig:eis_scheme}e). Note that the value $k_B T$ corresponds to a magnitude of a social noise or ``social temperature'' \cite{lider, castellano} in a proper Langevin equation.

\section{NUMERICAL SIMULATIONS}
\begin{figure}
\centering
\includegraphics[width=0.85\columnwidth]{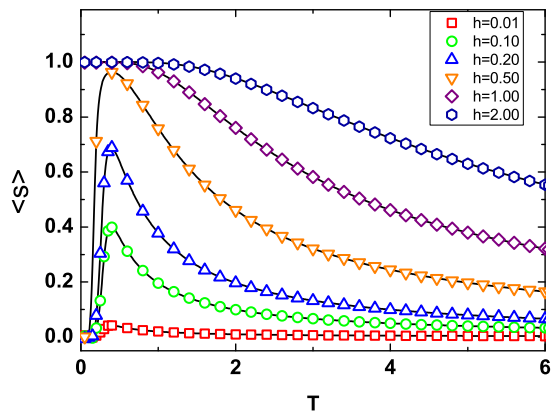}
\caption{(color online) Average spin $\langle s \rangle$ as function of temperature $T$ for different values of the external magnetic field (symbols) Solid lines come from Eqs (\ref{eq:e0}) and (\ref{eq:e1}). All data points are for $N=10^3$.}
\label{fig:eis_mT}
\end{figure}

Without losing the generality all numerical simulations have been performed for $J=k_B=1$. The average spin in the chain (an equivalent of the average emotion in online discussion) is calculated as $\langle s \rangle = \frac{1}{N} \sum_{n=1}^{N}s_n$ and afterwards averaged over $M$ realizations (typically, in this study $M=10^5$). Fig. \ref{fig:eis_mT} shows the average spin $\langle s \rangle$ as a function of the temperature $T$ for selected values of external field $h$. In the case of $h < 1$ the plot reveals $\langle s \rangle$ equal to zero for small $T \ll 1$, then a clear maximum for some specific crossover value $T_c$ appears. Finally, a decrease toward zero for $T > T_c$ takes place. In the case of $h \ge 1$ such a phenomenon is not observed: instead, $\langle s \rangle=1$ for small $T$ and then there is a monotonous decrease toward zero.

\begin{figure}
\centering
\includegraphics[width=0.85\columnwidth]{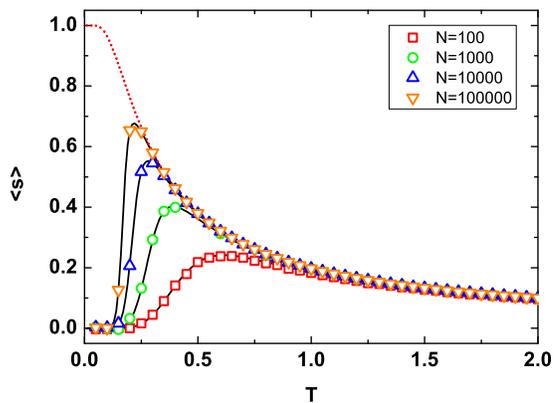}
\caption{(color online) Average spin $\langle s \rangle$ as function of temperature $T$ for different values of the chain size $N$ (symbols); all data points for $h=0.1$. Solid lines come from Eq. (\ref{eq:e0}) and the dotted line is $\tanh(2h/T)$.}
\label{fig:eis_eT}
\end{figure}

\begin{figure}
\centering
\includegraphics[width=\columnwidth]{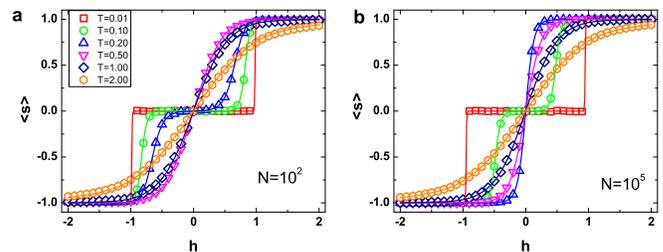}
\caption{(color online) Average spin $\langle s \rangle$ as function of the external field $h$ for different values of temperature $T$ (symbols). Solid lines come from Eq. (\ref{eq:e0}).}
\label{fig:eis_mh}
\end{figure}

Figure \ref{fig:eis_eT} shows that for smaller systems  (e.g., $N  = 10^2$) the crossover temperature $T_c$ is of order $0.5-1$ and is shifted toward lower values for larger systems. It is also interesting to track the dependence of the average spin value on the external field (see Fig. \ref{fig:eis_mh}). In the case of low temperatures ($T < T_c$) average spin value changes abruptly from $\langle s \rangle=-1$ to $\langle s \rangle=0$ for $h=-1$ and then from $\langle s \rangle=0$ to $\langle s \rangle=1$ for $h=1$. For higher temperatures ($T > T_c$) this change is smoother and length range of $h$ for which $\langle s \rangle \approx 0$ is smaller.

\section{ANALYTICAL DESCRIPTION} 
\begin{figure}
\centering
\includegraphics[width=\columnwidth]{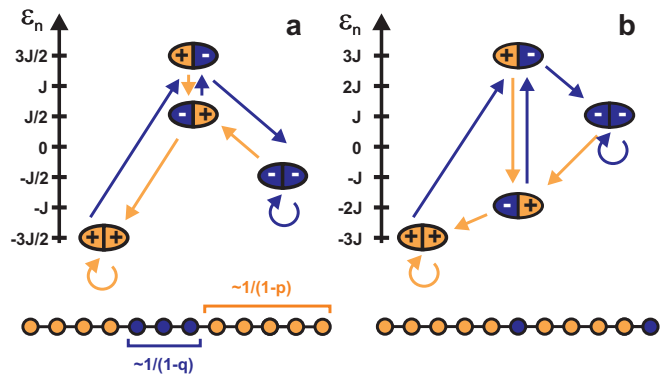}
\caption{(color online) Possible states of the \emph{last two} spins of the chain (the last is on the right of the pair) and the associated energies $\mathcal{E}_n$ tied to the last spin $n$. Arrows show how adding new + (orange) or - (blue) spin changes last two spin state (they are not the same spins after such step). Note that energy of starting state is of no importance, only relation between levels of possible destinations. The first situation ($0<h=J/2<J$) is bistable with two energy minima corresponding to states $s_{n-1}=s_{n}=+1$ and $s_{n-1}=s_{n}=-1$, while second $h=2>J$ is monostable.}
\label{fig:ks}
\end{figure}

The system dynamics can be easily described using a two-state Markov chain approach \cite{norris}. The growth of the chain follows the transition matrix $\mathbf{P}$
\begin{equation}\label{eq:p}
\mathbf{P}=\left[ \begin{array}{cc}
p & 1-p \\
1-q & q
\end{array} \right]
\end{equation}
with conditional probabilities $p = \Pr \left(+|+\right)$ and $q = \Pr \left(-|-\right)$. Matrix $\mathbf{P}$ defines probabilities evolution of both states $\Pr(s_{n}=\pm 1)$ as $\mathbf{s}_{n+1} = \mathbf{s}_{n}\mathbf{P}$, where $\mathbf{s}_{n} = \left[ \begin{array}{cc}\Pr(s_{n}=+1) & \Pr(s_{n}=-1)\end{array} \right]$. The average spin in the $n$-th node is $\langle s_n \rangle = \mathbf{s}_{0} \mathbf{P}^{n} \left[ \begin{array}{cc}1 & -1\\\end{array} \right]^{T}$ with $\mathbf{s}_0 = \left[ \begin{array}{cc}1/2 & 1/2\\\end{array} \right]$. Finally, the mean $\langle s_n \rangle$ calculated over all nodes in the chain equals to
\begin{equation}\label{eq:s}
\langle s \rangle =  \frac{p-q}{2-p-q} \left[ 1 + \frac{1}{N} - \frac{1 - (p+q-1)^{N+1}}{N(2-p-q)}\right],
\end{equation}
The specific values of $p$ and $q$ for our model are
\begin{equation}\label{eq:hsmall}
\left\{
\begin{array}{l}
p = \Pr \left(+|+\right) = 1-\frac{1}{2} \mathrm{e}^{-\widetilde{\beta}(h+J)}\\
q = \Pr \left(-|-\right) = \frac{1}{2} \pm \frac{1}{2} \mp \frac{1}{2} \mathrm{e}^{\pm \widetilde{\beta}(h-J)}\\
\end{array}
\right.,
\end{equation}
where upper signs correspond to case $|h| < J$, lower signs to $|h| \ge J$ and $\tilde\beta=2/(k_B T)$ (see Appendix \ref{appa} for details).
In further discussion we assume that $h>0$, although all derivations and effects are also true for $h<0$ with reversed spins. Different form of $q$ for small and large $|h|$ follows from the interchange of energy level positions corresponding to states $s_n=s_{n+1}=-1$ and $s_n=-1,s_{n+1}=+1$ (see Fig. \ref{fig:ks} and Appendix \ref{appa}).
Putting (\ref{eq:hsmall}) into (\ref{eq:s}) we get the average spin in the chain  for low magnetic fields $|h| < J$ as:
\begin{equation}\label{eq:e0}
\langle s \rangle_s = \tanh  \widetilde{\beta} h \left[1 + \frac{1}{N} - \frac{1 - \left(1- \mathrm{e}^{- \widetilde{\beta} J} \cosh  \widetilde{\beta} h \right)^{N+1}}{N \mathrm{e}^{- \widetilde{\beta} J} \cosh  \widetilde{\beta} h} \right]
\end{equation}
and for $|h| \ge J$ as:
\begin{equation}\label{eq:e1}
\langle s \rangle_l = \frac{\cosh \widetilde{\beta} J - \mathrm{e}^{ \widetilde{\beta} h}}{\sinh  \widetilde{\beta} J - \mathrm{e}^{ \widetilde{\beta} h}} \left[1 + \frac{1}{N} - \frac{1 - \left( \mathrm{e}^{- \widetilde{\beta} h} \sinh  \widetilde{\beta} J \right)^{N+1}}{N \left(1 - \mathrm{e}^{- \widetilde{\beta} h} \sinh  \widetilde{\beta} J \right)} \right]
\end{equation}
Let us note that factors standing in front of square brackets of Eqs. (\ref{eq:e0}) and (\ref{eq:e1}) describe the thermodynamical limit and coincide with the corresponding factor in Eq. (\ref{eq:s}). These analytical results are fully supported by numerical simulations (solid lines in Figs \ref{fig:eis_mT}, \ref{fig:eis_eT} and \ref{fig:eis_mh}). 

\begin{figure}
\centering
\includegraphics[width=0.85\columnwidth]{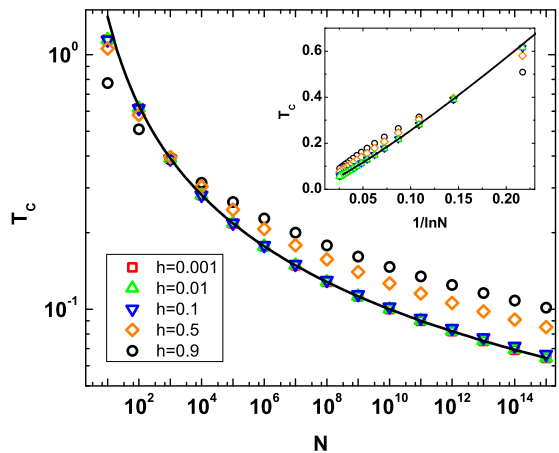}
\caption{(color online) Crossover temperature $T_c$ versus the size of the chain $N$. Symbols are numerical solution of Eq. (\ref{eq:e0}) while the solid line comes from Eq. (\ref{eq:tc}). Note that decay of $T_c$ is very slow. The symbols for three lowest $h$ values overlap. The inset shows $T_c$ versus $1/\ln N$.}
\label{fig:eis_size}
\end{figure}

Both numerical and analytical approaches indicate crucial role played by the system's size $N$ (see Fig. \ref{fig:eis_size}): for a constant value of external field $h$ increasing $N$ leads to a shift in $T_c$ toward $T=0$ as well as to an increase of the maximum value $\langle s \rangle(T_c)$. To get an analytical estimation of $T_c$ we assume that $\widetilde{\beta}h \ll 1$ which gives the opportunity to rewrite Eq. (\ref{eq:e0}) as
\begin{equation}\label{eq:ea}
\langle s \rangle_s \approx \widetilde{\beta} h \left[ 1 + \frac{1}{N} -\frac{1-\left(1-\mathrm{e}^{-\widetilde{\beta}J}\right)^{N+1}}{N \mathrm{e}^{-\widetilde{\beta}J}} \right]
\end{equation}Because it is linearly dependent on $h$, the factor is an equivalent of the susceptibility $\left( \frac{\partial \langle s \rangle_s}{\partial h} \right)_{h=0}$ times $h$. If we further assume $N \gg 1$, and solve $\frac{\partial \langle s \rangle_s}{\partial T}=0$ one can approximate $T_c$ as
\begin{equation}\label{eq:tc}
T_c \approx \frac{2 J}{k_B \left[\mathrm{W}(N\mathrm{e})-1 \right]},
\end{equation}
where $\mathrm{W}(...)$ is Lambert $\mathrm{W}$ function. Comparison between this approximation and numerical solution of Eq. (\ref{eq:e0}) is shown in Fig. \ref{fig:eis_size}, providing evidence of good agreement for small values of $h$ as expected.  

\section{PHENOMENOLOGICAL DESCRIPTION}
The striking difference between average spin values for low and high fields $h$ - as  presented by Eqs. \ref{eq:e0} and \ref{eq:e1} - can be explained as follows. For $0 < h < J$ (Fig. \ref{fig:ks}a) the four states system of two last spins $s_{n-1}$ and $s_{n}$   possess two lowest energy states  corresponding to parallel ordering of both spins   $s_{n-1}=s_{n}=-1$  or $s_{n-1}=s_{n}=+1$. Such a system is {\it bistable} and temperature causes a random switching between clusters (domains) of opposite spin values. The average length $l_{+}$ of a spin up cluster is $l_{+} = 1/(1-p)$ while corresponding length $l_{-}$ of spin down cluster is $l_{-} = 1/(1-q)$. Note that in the thermodynamical limit Eq. \ref{eq:s} can be written as $\langle s \rangle =(\frac{l_{+}}{l_{-}}-1)/(\frac{l_{+}}{l_{-}}+1)$. The quotient $l_{+}/l_{-}$ for $|h| < J$ is equal to $\mathrm{e}^{2 \widetilde{\beta}h}$ thus it is independent from $J$. Of course with increasing $J$  lengths $l_{+}$, $l_{-}$ of both types of clusters increase but it does not influence the mean spin $\langle s \rangle$ of the {\it infinite} chain.
After crossing the critical value of the magnetic field $h=J>0$ the situation changes. The energy $\mathcal{E}_n$ for $s_{n-1}=s_{n}=-1$  is higher than the energy $\mathcal{E}_n$ for $s_{n-1}=-1, $ $s_{n}=+1$, making system {\it monostable} - thus the temperature mostly causes {\it single} spins $s_{n}=-1$ to appear in the chain dominated by the stable $s_{n}=+1$ phase. It means that there are no clusters of negative spins (Fig. \ref{fig:ks}b) for  $|h| > J$ and the mean magnetization depends mainly on the density of single spin impurities. In fact, for $h=J$ we have $l_{-}=2$ and $l_{-}$ further decays with the increase of $h$. However, the density of single spin impurities is a decreasing function of an energy of interface between $s_{n-1}=+1$ and  $s_{n}=-1$ that is dependent on the coupling constant $J$. This leads to a profound difference between mean values of spins in the chain in the case of the thermodynamical limit of Eq. (\ref{eq:e0}) and Eq. (\ref{eq:e1}). The first one takes the form of $\langle s \rangle_s = \tanh \widetilde{\beta}h$ which is independent of $J$. In fact, for a chain of a finite length $N$ there is always an influence of the boundary condition that leads to the emergence of the first (boundary) cluster with spins up or spins down. Due to the assumed symmetry $\Pr(s_0= \pm 1)=1/2$ both types of these boundary clusters are represented with the same probability.
It follows that a short chain possesses a zero mean magnetization when one averages it over an ensemble of initial/boundary conditions what can be observed at Fig. \ref{fig:eis_eT}. While increasing the length $N$ the chain magnetization depends more and more on the ratio between lengths $l_{+}/l_{-}$ of positive and negative clusters.
It follows that the effect of the boundary condition disappears the faster the smaller is the coupling constant $J$ responsible for spin clustering. In the thermodynamical limit the presence of the boundary cluster can be disregarded and the magnetization does not depend on the coupling $J$. This phenomenological picture can also justify a non-monotonous dependence of mean magnetization on the temperature when  $|h| < J$. When the temperature is low the length of initial cluster tends to infinity, thus mean magnetization can be close to zero even for large systems because of a random, symmetric initial condition. If the temperature increases both lengths $l_{+}$,$l_{-}$ decay and thus the effect of boundary conditions becomes less important and the mean magnetization increases toward magnetization of infinite chain governed by $l_{+}/l_{-}$. However, the ratio $l_{+}/l_{-}$ decreases with $T$, thus for higher temperatures $\langle s \rangle$ decreases toward $0$. It follows there is a {\it crossover temperature} $T_c$ where the magnetization is maximal in the effect of interplay between the initial condition and temperature fluctuations (see Fig. \ref{fig:eis_eT}). This crossover temperature decays with system size $N$ (see Fig. \ref{fig:eis_size}), since for larger systems the impact of the first cluster is very small. Let us note that for $|h| > J$, when no clusters are present in the system there is no crossover temperature and the magnetization in the thermodynamical limit depends on the coupling constant $J$: $\langle s \rangle_l \approx (\cosh \widetilde{\beta} J - \mathrm{e}^{ \widetilde{\beta} h})/(\sinh  \widetilde{\beta} J - \mathrm{e}^{ \widetilde{\beta} h})$.

\section{COMPARISON WITH CLASSICAL 1D ISING MODEL}
\begin{figure}
\centering
\includegraphics[width=\columnwidth]{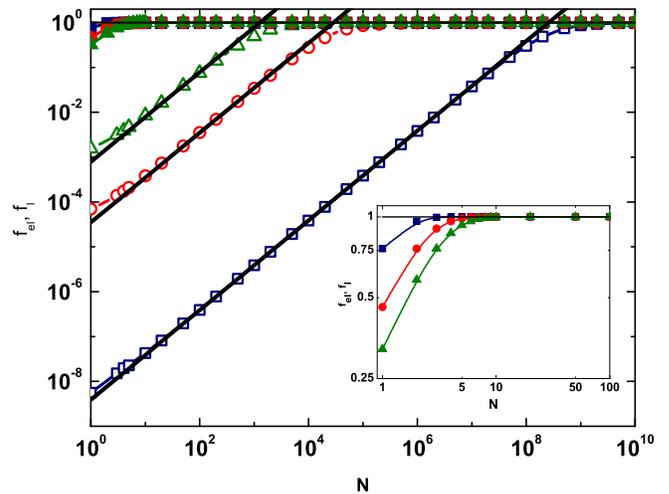}
\caption{(color online) Factors $f$ (for our model, empty symbols) and $f_I$ (for the classical Ising model, filled symbols) versus the size of the chain $N$ for $h=0.1$. Symbols (squares for $T=0.1$, circles $T=0.2$, triangles for $T=0.3$) come from Eqs (\ref{eq:fg}-\ref{eq:fi}) while straight lines are Eq. (\ref{eq:fga}). The inset magnifies the results for the Ising model for $N \in [1;100]$.}
\label{fig:f}
\end{figure}

It is of use to compare and contrast the obtained results with the classical one-dimensional Ising model (e.g., \cite{statmech}). The most noticeable difference undoubtedly regards the foundations of the model --- in the classical case the length of the chain is fixed and each node $n$ is initially filled with a spin $s_n = \pm 1$. All spins can repeatedly change in time and their dynamics involve energy coupling with both neighbors. In our model, the chain grows, and only the newest spin added is subject to dynamics for a single time step, taking only its predecessor into account, after which it is quenched and unchanging. This equates system size $N$ to time. The parameter $N$ plays a pivotal role governing the position and height of the maximum of  $\langle s \rangle$  for $h < J$. 

The magnetization per spin for $J > 0$ (ferromagnetic case) --- an equivalent of $\langle s \rangle$, in the case of the classical one-dimensional Ising model is given by 
\begin{equation}\label{eq:mi}
m(h,T,N) = \frac{\sinh \beta h}{\sqrt{\sinh^2\beta h + \mathrm{e}^{-4 \beta J}}} \frac{1 - \left( \frac{\lambda_-}{\lambda_+} \right)^N}{1 + \left( \frac{\lambda_-}{\lambda_+} \right)^N}
\end{equation}
where
\begin{equation}\label{eq:eigen}
\lambda_{\pm} = \mathrm{e}^{\frac{1}{T}} \left( \cosh \beta h \pm \sqrt{\mathrm{e}^{-4 \beta J} + \sinh^2 \beta h} \right) 
\end{equation}
are the eigenvalues of the transfer matrix \cite{statmech}. The magnetization is a strictly monotonous decaying function of $T$ starting from $m=1$ for $h > 0$ and it rapidly converges with system size $N$ to its asymptotic value. On the other hand, from Eq. (\ref{eq:tc}) it can be concluded that for our model, this convergence is much slower. The dependence of $T_c$ is even slower, as $1/\ln N$ [see Eq. (\ref{eq:tc}) and Fig. \ref{fig:eis_size}]. A comparison of the influence of the chain size in both models is presented in Fig. \ref{fig:f} where 
\begin{equation}\label{eq:fg}
f =  1 + \frac{1}{N} - \frac{1 - \left(1- \mathrm{e}^{- \widetilde{\beta} J} \cosh  \widetilde{\beta} h \right)^{N+1}}{N \mathrm{e}^{- \widetilde{\beta} J} \cosh  \widetilde{\beta} h}
\end{equation}
and 
\begin{equation}\label{eq:fi}
f_I = \frac{1 - \left( \frac{\lambda_-}{\lambda_+} \right)^N}{1 + \left( \frac{\lambda_-}{\lambda_+} \right)^N}
\end{equation}
are the factors in, respectively, $\langle s \rangle_s$ [see Eq. (\ref{eq:e0})] and $m(h,T,N)$ [see Eq. (\ref{eq:mi})], that are dependent on the chain size $N$. Factor $f_I$ quickly converges to 1, e.g., for $J=k_B=1$, $h=0.1$ and $T=0.1$ one needs as little as $N_c=10$ to have $|f_I(N \rightarrow \infty) - f_I(N_c)| < 0.001$, while the factor $f$, depending on the value of $T$ and $h$, can need a large chain length in order to reach the thermodynamic limit. In fact, for small $T$ Eq. (\ref{eq:fg}) can be approximated by 
\begin{equation}\label{eq:fga}
f \approx \frac{1}{2} (N + 1) \mathrm{e}^{-\widetilde{\beta} J}\cosh \widetilde{\beta}h, 
\end{equation}
shown as straight lines in Fig.  \ref{fig:f}, which in turn can be used to estimate the critical value of $N$ for which $f$ is equal to one as $N_c \approx 4 \mathrm{e}^{\widetilde{\beta} (J - h)}$. Thus, for $J=k_B=1$, $T=0.1$ and $h=0.1$ we get $N_c \approx 2.6 \times 10^8$.  

\section{COMPLETE THERMALIZATION OF INDIVIDUAL SPINS}
\begin{figure}
\centering
\includegraphics[width=\columnwidth]{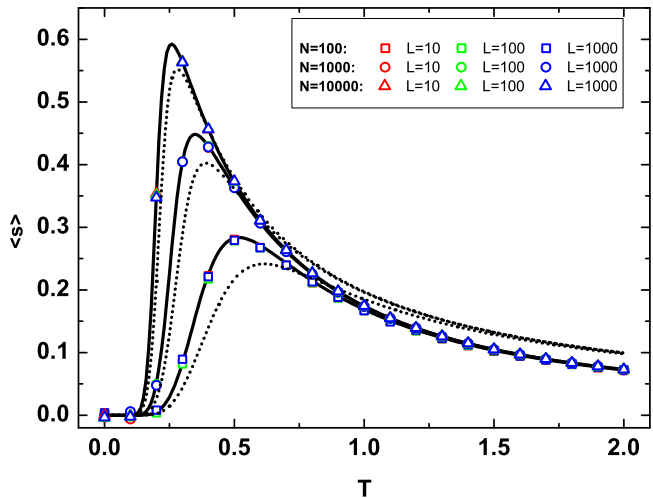}
\caption{(color online) Chain magnetization $\langle s \rangle$ versus temperature $T$ for single-step update [dotted line, given by Eq. (\ref{eq:e0})] and single spin thermalization [solid line, given by Eq. (\ref{eq:sst})] with $h=0.1$ and $J=1$. Points represent numerical simulations for chain of length $N$ with $L$ steps of update procedure averaged over $M=10^5$ realizations.}
\label{fig:b}
\end{figure}

If instead of performing a single Metropolis update of the newly attached spin we allow it to fully thermalize, then our newly added spin $n$ is essentially drawn from the canonical ensemble. Therefore the probabilities $p$ and $q$ can be derived by using Boltzmann factors.
The probability $p=\Pr(+|+)$ is
\begin{equation}
p = \frac{\mathrm{e}^{-\beta \mathcal{E}_n(+1)}}{\mathrm{e}^{-\beta \mathcal{E}_n(+1)} + \mathrm{e}^{-\beta \mathcal{E}_n(-1)}}
\end{equation}
where $\mathcal{E}_n(s_n) = -J s_{n-1} s_n - h s_n$.
If we put it into our formula, we obtain
\begin{equation}
p = \Pr(+|+) = \frac{\mathrm{e}^{\beta(J+h)}}{\mathrm{e}^{\beta(J+h)} + \mathrm{e}^{-\beta(J+h)}}
\end{equation}
which further implies that
\begin{equation}
1-p = \Pr(-|+) = \frac{\mathrm{e}^{-\beta(J+h)}}{\mathrm{e}^{\beta(J+h)} + \mathrm{e}^{-\beta(J+h)}}
\end{equation}
Similarly, the probabilities $q$ and $1-q$ can be written as
\begin{eqnarray}
q = \Pr(-|-) = \frac{\mathrm{e}^{\beta(J-h)}}{\mathrm{e}^{\beta(J-h)} + \mathrm{e}^{-\beta(J-h)}}\\
1-q = \Pr(+|-) = \frac{\mathrm{e}^{-\beta(J-h)}}{\mathrm{e}^{\beta(J-h)} + \mathrm{e}^{-\beta(J-h)}}
\end{eqnarray}
Using the Markov chain approach [Eq. \ref{eq:s}], we can determine the mean spin $\left\langle s \right\rangle$ and finally write it as
\begin{equation}\label{eq:sst}
\langle s \rangle = \frac{\sinh \widetilde{\beta}h}{\cosh \widetilde{\beta}h + \mathrm{e}^{-\widetilde{\beta}J}} \left[1+ \frac{1}{N} - \frac{1-u^{N+1}}{N(1-u)} \right],
\end{equation}
where
\begin{equation}
u = \frac{\sinh \widetilde{\beta} J}{2 \cosh \frac{\widetilde{\beta}(h+J)}{2} \cosh \frac{\widetilde{\beta}(h-J)}{2}}
\end{equation}
The behavior of the model with spin thermalization, while somewhat different quantitatively from the single-update approach, is still qualitatively the same, exhibiting the maximum of $\langle s \rangle(T)$. One notable difference is the absence of the threshold $h=J$ where the probabilities $p$ and $q$ change their forms, and subsequently a fully smooth transition between $h<J$ and $h>J$ regimes. Figure \ref{fig:b} presents a comparison between single update (solid line) and spin thermalization approaches (dotted line) for $h < J$ supported with numerical simulations. The plot proves that although there is a difference in the crossover temperature as well as in the peak height the character of the curve is kept the same.

\section{REAL DATA}
The developed method bears some perspectives for possible comparison with the real data. The current model could be generalized for unequal probabilities $\Pr(s_0=+1)=1-\Pr(s_0=-1)$ and if the conditional probabilities $p$ and $q$ were known from the real data (as in the case presented in \cite{plos}), then properly modified Eq. (\ref{eq:hsmall}) might be used for obtaining values of $h$ and $T$. However it is essential to notice that in fact the probabilities $\Pr(s_0=+1)$ and $\Pr(s_0=-1)$ are unknown as they are {\it a priori} values. The other difficulty comes from the fact, that in the real-data study \cite{plos} the conditional probability is calculated using all data while for comparison purposes it should be done for each value of $N$ separately. Nonetheless the sketched procedure is possible to be accomplished.

\section{CONCLUSIONS}
In summary we have demonstrated that the finite system size and initial conditions can lead to the emergence of a non-monotonous dependence of the mean magnetization on system's temperature in a growing one-dimensional Ising model with quenched spins. The effect exists only for magnetic fields smaller than the value of the spin coupling constant and the crossover temperature decays to zero very slowly with the system size. Using Markov chain theory we have developed an analytical approach to this phenomenon that well fits numerical simulations. The effect can be understood as a competition between thermal fluctuations and the influence of the initial condition that  fixes  orientation of spins in the first cluster. The crossover temperature can be explained as the point where the initial ordered cluster (domain) is no longer dominant thanks to thermal fluctuations, yet the temperature did not lower much the average magnetization toward zero. The effect exists also when  spins are not quenched but fully thermalized after the attachment to the chain. The absence of the effect for the higher magnetic field is the result of a transition from a bistable to a monostable  energy landscape of a pair of neighboring spins. We think that, although directly inspired by the clustering phenomena observed in the online emotional discussions \cite{plos,ania}, the model can open interesting playground for all systems where initial conditions and finite size effects are relevant.

\begin{acknowledgements}
This work was supported by a European Union grant by the 7th Framework Programme, Theme 3: Science of Complex Systems for Socially Intelligent ICT. It is part of the CyberEMOTIONS (Collective Emotions in Cyberspace) project (contract 231323). We also acknowledge support from Polish Ministry of Science Grant 1029/7.PR UE/2009/7.
\end{acknowledgements}

\appendix
\section{DERIVATION OF THE MEAN MAGNETIZATION}\label{appa}
Let us calculate the probability that a spin-up follows another spin-up. We assume the presence of external field $h \ge 0$. First, we set $s_0=+1$. Then, with equal probabilities $\Pr(s_1 \pm 1)=1/2$, spin in the next node is chosen to be up or down. Next, we calculate the change of function $\mathcal{E}$, given by 
\begin{equation}
\Delta \mathcal{E} = \mathcal{E}'_1 - \mathcal{E}_1 = - (Js_0 + h)(s'_1-s_1)
\end{equation}
that follows 
\begin{enumerate}
\item if $s_1=+1$ and $s'_1=-1$ then $\Delta \mathcal{E}=2(h+J)>0$, so the change is accepted with probability equal to $\mathrm{e}^{-\widetilde{\beta}(h+J)}$ and not accepted with probability equal to $1 - \mathrm{e}^{-\widetilde{\beta}(h+J)}$, where $\widetilde{\beta} = 2/(k_B T)$, 
\item if $s_1=-1$ and $s'_1=+1$ then $\Delta \mathcal{E}=-2(h+J)<0$, so the change is always accepted. 
\end{enumerate}
As a consequence the probability $p_{++}$ of a spin-up following another spin-up is equal to $p_{++} = \frac{1}{2}(1 - \mathrm{e}^{-\widetilde{\beta}(h+J)}) + \frac{1}{2} \times 1 = 1-\frac{1}{2}\mathrm{e}^{-\widetilde{\beta}(h+J)}$. Then, for $h \ge 0$ we have
\begin{equation}\label{eq:h1a}
\left\{
\begin{array}{l}
p_{++} = \Pr \left(+|+\right) = 1-\frac{1}{2} \mathrm{e}^{-\widetilde{\beta}(h+J)}\\
p_{+-} = \Pr \left(-|+ \right) = 1 - p_{++}\\
\end{array}
\right.
\end{equation}

\begin{figure*}[ht]
\centering
\includegraphics[width=0.8\textwidth]{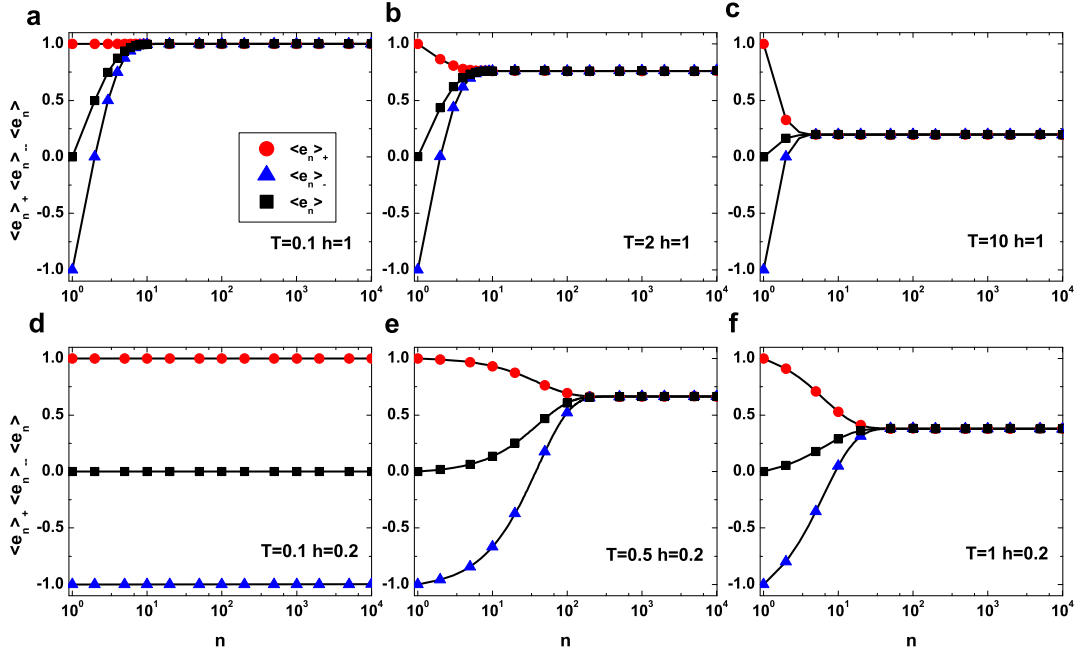}
\caption{(color online) Average spin $\langle s_n \rangle$ in the $n$-th node and average spin if the $n$-th node starting from an up spin $\langle s_n \rangle_{+}$ and down spin $\langle s_n \rangle_{-}$ for selected parameters $h$ and $T$. Symbols are numerical simulations and solid lines come from Eqs. (\ref{eq:enp}), (\ref{eq:enm}) and (\ref{eq:enpqa}).}
\label{fig:eis_en}
\end{figure*}

Now let us calculate the probability that a spin-down follows another spin-down. Contrary to the previous case we set $s_0=-1$ and then, with equal probabilities $\Pr(s_1 \pm 1)=1/2$, spin in the next node is chosen to be up or down. Next, we calculate the change of function $\mathcal{E}$ 
\begin{enumerate}
\item if $s_1=-1$ and $s'_1=+1$ then $\Delta \mathcal{E}=-2(h-J)$, 
\item if $s_1=+1$ and $s'_1=-1$ then $\Delta \mathcal{E}=2(h-J)$. 
\end{enumerate}
Here, the issue of the spin change being accepted or not depends on the value of external field:
\begin{itemize}
\item if $0 \le h < J$ then
	\begin{enumerate} 
		\item for $s_1=-1$ and $s'_1=+1$ we have $\Delta \mathcal{E} > 0$ so the change is accepted with probability $\mathrm{e}^{\widetilde{\beta}(h-J)}$ and not accepted with probability equal to $1-\mathrm{e}^{\widetilde{\beta}(h-J)}$ 
		\item for $s_1=+1$ and $s'_1=-1$ we have $\Delta \mathcal{E} < 0$ so the change is always accepted, 
	\end{enumerate}
\item if $h \ge J$ then the character $-1 \rightarrow +1$ and $+1 \rightarrow -1$ changes since signs of energy difference $\Delta \mathcal{E}$ rearrange
	\begin{enumerate} 
		\item for $s_1=-1$ and $s'_1=+1$ we have $\Delta \mathcal{E} < 0$ so the change is always accepted, 
		\item for $s_1=+1$ and $s'_1=-1$ we have $\Delta \mathcal{E} > 0$ so the change is accepted with probability $\mathrm{e}^{-\widetilde{\beta}(h-J)}$ and not accepted with probability equal to $1-\mathrm{e}^{-\widetilde{\beta}(h-J)}$. 
	\end{enumerate}
\end{itemize}

As a consequence the probability $p_{--}$ of a spin-down following another spin-down is equal to $p_{--} = \frac{1}{2}(1 - \mathrm{e}^{\widetilde{\beta}(h-J)}) + \frac{1}{2} \times 1$ for $0 \le h < J$ and $p_{--} = \frac{1}{2} \times 0 + \frac{1}{2} \times \mathrm{e}^{-\widetilde{\beta}(h-J)}$ for $h \ge J$. Thus we have
\begin{equation}
\left\{
\begin{array}{l}\label{eq:h2a}
p_{--} = \Pr \left(-|-\right) = 1-\frac{1}{2} \mathrm{e}^{\widetilde{\beta}(h-J)}\\
p_{-+} = \Pr \left(+|- \right) = 1 - p_{--}
\end{array}
\right.
\end{equation}
for $0 \le h < J$ and
\begin{equation}
\left\{
\begin{array}{l}\label{eq:h3a}
p_{--} = \Pr \left(-|-\right) = \frac{1}{2} \mathrm{e}^{-\widetilde{\beta}(h-J)}\\
p_{-+} = \Pr \left(+|- \right) = 1 - p_{--}
\end{array}
\right.
\end{equation}
for $h \ge J$.

For simplicity of the further calculations we use the following notation: the probability to stay in $+1$ state is $p_{++}=p$, the probability to move from state $+1$ to state $-1$ is $p_{+-}=1-p$. Similarly we denote the probability to stay in $-1$ as $p_{- -}=q$ and the probability to move from state $-1$ to $+1$ is $p_{- +}=1-q$. In effect we obtain the transition matrix $\mathbf{P}$ given by Eq. (\ref{eq:p}), that defines probabilities evolution of both  states $\Pr(s_{n}=\pm 1)$ 
\begin{equation}
\mathbf{s}_{n+1} = \mathbf{s}_{n}\mathbf{P},
\end{equation}
where $\mathbf{s}_{n} = \left[ \begin{array}{cc}\Pr(s_{n}=+1) & \Pr(s_{n}=-1)\\\end{array} \right]$. Thus the evolution of $\mathbf{s}_{n}$ is in fact an equivalent of a two-state Markov chain \cite{norris} governed by the transition matrix $\mathbf{P}$. Appropriate elements of the $n$-th power of matrix $\mathbf{P}$ give the probabilities that the chain that started with a specific spin has a certain spin in its $n$-th node (e.g., $(\mathbf{P}^n)_{11}$ is the probability that after starting $s_0=+1$ the spin in the $n$-th node will also be also $s_n=+1$). A short algebra leads to
\begin{equation}
\mathbf{P}^n=\left[ \begin{array}{cc}
\frac{q-1+(p-1)(q+p-1)^n}{q+p-2} & \frac{(p-1)[1-(q+p-1)^n]}{q+p-2} \\
\frac{(q-1)[1-(q+p-1)^n]}{q+p-2} & \frac{p-1+(q-1)(q+p-1)^n}{q+p-2} \\
\end{array} \right]
\end{equation}

Subtracting the second column from the first one in matrix $\mathbf{P}^n$ leads to equations describing the average spin values $\langle s^n \rangle_{\pm}$ in the $n$-th node assuming that the first node contained a specific spin orientation ($s_0=+1$ or $e_0=-1$):
\begin{eqnarray}
(\mathbf{P}^n)_{11} - (\mathbf{P}^n)_{12} = \langle s_n \rangle_{+} = \frac{p-q-2(p-1)(p+q-1)^n}{2-p-q}\label{eq:enp}\\
(\mathbf{P}^n)_{21} - (\mathbf{P}^n)_{22} = \langle s_n \rangle_{-} = \frac{p-q+2(q-1)(p+q-1)^n}{2-p-q}\label{eq:enm}.
\end{eqnarray}
Calculating the average value of $\langle s_n \rangle_{+}$ and $\langle s_n \rangle_{-}$ leads to the average spin in the $n$-th node
\begin{equation}\label{eq:enpqa}
\langle s_n \rangle = \frac{\langle s_n \rangle_{+} + \langle s_n \rangle_{-}}{2} =  \frac{(p-q)[1-(p+q-1)^n]}{2-p-q},
\end{equation}
The plots of $\langle s_n \rangle_{+}$, $\langle s_n \rangle_{-}$ and $\langle s_n \rangle$ versus $n$ for selected values of the external field $h$ and temperature $T$ are shown in Fig. \ref{fig:eis_en}. One can easily observe the convergence of $\langle s_n \rangle$ to a constant value for a sufficiently large value of $n$. In fact, as $p + q - 1 < 1$ we have $\lim \limits_{n \to \infty} \langle s_n \rangle = (q-p)/(p+q-2)$.

Finally, performing the sum of $\langle s_n \rangle$ over all nodes in the chain gives the average spin:
\begin{equation} \label{eq:avgsa}
\langle s \rangle = \frac{1}{N} \sum_{n=1}^{N} \langle s_n \rangle.
\end{equation}

Similar calculations can be performed for ranges $h \in (-\infty; -J]$ and $h \in (-J;0]$. The symmetry of the problem results in swapping all the indices "+" to "-" and likewise in Eqs (\ref{eq:h1a}-\ref{eq:h3a}). As an outcome we obtain a rotated matrix $\mathbf{P}$ that leads again to Eq. (\ref{eq:enpqa}). In effect by applying exact values of the probabilities $p$ and $q$ given by Eqs (\ref{eq:h1a}-\ref{eq:h3a}) we obtain the average spin for $|h| < J$ as (\ref{eq:e0}) and for $|h| \ge J$ as
(\ref{eq:e1}).

\end{document}